\documentclass[twocolumn,twoside]{cap-journal}

\usepackage{url}

\newtheorem{definition}{Definition}
\newtheorem{theorem}{Theorem}
\newtheorem{proposition}{Proposition}
\newtheorem{corollary}{Corollary}

\makeatletter
\renewcommand{\author}[1]{%
    \global\advance\@authcnt1
    \expandafter\gdef\csname @auth\romannumeral\the\@authcnt\endcsname
    {\begin{tabular}[t]{@{}c@{}}\twlsfb #1\\[20pt]
     \end{tabular}}}
\makeatother

\title{AN ENGINEERING READING OF THE BOUNDED\\
HALTING PROBLEM AND OBSERVABLE PROPERTIES\\
OF A UNIVERSAL MACHINE}

\author{\twlsfb Sergey Salishev
    \affiliation{
      St. Petersburg State University\\
      Saint Petersburg\\
      199034\\
      Russia\\
      s.salischev@spbu.ru
    }
}

\begin{document}

\maketitle

\begin{abstract}
The halting problem asks whether a program eventually halts, with
unrestricted resources. Engineering systems instead certify bounded
halting: given a program, an input, and a step budget T, decide whether
the program halts within T steps. This paper studies two of its uses.
First, it defines an exact finite-budget problem,
placing the simulated program and the checker on the same unit-cost
timescale, so the checker can be embedded in an adversarial program. A
diagonal argument with explicit additive overhead shows that no fixed
deterministic checker decides every instance within any fixed fraction
of T below one, for large T. In the nondeterministic case, T bounds
the branch depth allowed for a stable observable decision. Second,
bounded-halting instances give adversarial tests for finitary
universality, since a universal model must support arbitrary finite
computations as T grows. The tests check resource growth with T, the
absence of a fixed deliberation bound, and later access to intermediate
data. They separate a fixed model with extendable external resources, a
model family that grows with T, and a larger system built around a
fixed component. For machine-learning systems, this identifies the
resource growth a universality claim assumes and whether it fails for
deployed models.
\end{abstract}

\begin{keywords}
bounded halting problem, universal machine, nondeterminism, finite
resources, observable properties, machine-learning systems
\end{keywords}

\section{Introduction}
\label{sec:introduction}

Engineering systems operate with finite resource limits. A monitor
checks whether a run reaches a bad state before a deadline. A model
checker or symbolic executor often has a fixed search depth
\cite{biere1999symbolic,king1976symbolic,cadar2013symbolic}. A runtime
controller has limited memory, precision, and time. A machine-learning
component has a finite context and a bounded inference or generation
budget. The required result is an exact answer for a stated system
boundary and resource budget \(T\).

Bounded halting is the smallest formal version of that contract because
a program must halt before returning an answer. So checking it is a
practical test for correctness. Given a program, an
input, and a budget \(T\), it asks whether the program halts within
\(T\) steps. Direct simulation may require all \(T\) steps.
The engineering question is whether one fixed implementation can decide
every instance exactly in \(\alpha T\) steps for some fixed
\(\alpha<1\). This \(\alpha\) is the practically desired certification 
speedup over simulation. Theorem~\ref{thm:det-bound} and 
Theorem~\ref{thm:nondet-stable} show there is no such speedup in both 
deterministic and nondeterministic cases.

Although universality may seem unimportant for fixed resources, it comes
into play when the extensibility of the deployed system and the
trade-off between time and computation are considered.

The undecidability of unbounded halting is a universality
test~\cite{davis1958computability}.
A universal model must support arbitrary finite
simulations as the budget \(T\) grows. An exact bounded-halting instance is 
a stress test for the implementation. Thus,
the same finite problem connects exact budgeted decision with
certification of universal computation. The universal model should not have:
\begin{enumerate}
    \item a fixed state set size independent of \(T\);
    \item a fixed deliberation bound independent of \(T\);
    \item a fixed forgetting horizon independent of \(T\).
\end{enumerate}

To be universal, a fixed computation model may permit extendable external resources,
such as tape or stack, without changing its transition rules. A growing model family
may change its parameter count, precision, architecture, or exactly
addressable context range. A larger system may build up resources
around a fixed component.

Large language models give a current motivation for this distinction.
The Transformer architecture introduced in \cite{vaswani2017attention}
is an example of a modern system. A practical implementation has a fixed
parameter count, finite precision, bounded runtime, and a usable context
limited by its positional encoding accuracy. Universality results
commonly use a growing model family in which one or more of these
resources increase with \(T\) \cite{perez2021attention,merrill2023expressive},
giving a misleading impression for a single fixed model. A larger system can
also add storage, tool calls, or scheduler time around a fixed
component. Examples include retrieval-augmented generation
\cite{lewis2020retrieval} and multi-agent LLM systems
\cite{wu2024autogen}. In these cases, certification applies to the
combined system.

The unbounded formulation of the halting problem is fundamental for
computability theory.
But the halting behavior of a closed engineering system with finitely many
configurations is decidable in principle by cycle
detection~(Proposition~\ref{prop:finite-halting}). Its
configuration graph may still be too large to search within the
requested budget. Section~\ref{sec:related} reviews classical
undecidability, hierarchy, and speedup results relative to bounded halting.

The contributions are as follows:
\begin{enumerate}

\item
Theorem~\ref{thm:det-bound} states the deterministic fixed-budget
diagonal bound with explicit additive overhead, and
Corollary~\ref{cor:no-factor-shortcut} extracts the implementation
consequence. For any fixed \(\alpha<1\), exact uniform decision of
\(BH_T\) cannot run in \(\alpha T\) time on all inputs once \(T\) is
large enough.

\item Theorem~\ref{thm:nondet-stable} gives the nondeterministic reading:
branching supplies width, while a stable exact decision inherits the
branch-depth bound on deterministic subinstances. The same result is
recorded for finite probability-one randomized implementations in
Corollary~\ref{cor:prob-one}.

\item The paper gives implementation-level tests for universality.
Proposition~\ref{prop:finite-halting} prohibits fixed-size models.
Proposition~\ref{prop:deliberation} rules out fixed bounded deliberation,
and Proposition~\ref{prop:forgetting} records finite forgetting as an
obstruction to universal computation.

\item Table~\ref{tab:machine-classes} applies observable tests to standard
automata, multi-agent systems, and finite-resource Transformer variants,
showing that deployed individual Transformer models do not support
universal computation.

\end{enumerate}

\section{Preliminaries}
\label{sec:preliminaries}

Standard deterministic and nondeterministic Turing machines over finite
alphabets are used throughout \cite{hopcroft2001introduction}. A configuration
contains the finite control state, the tape contents that have been
reached so far, and the head positions. One transition has unit cost.
The exact choice of one-tape or multi-tape model only changes fixed or
standard simulation overheads; the diagonal statements below state the
overhead assumption explicitly.

Let \(U\) be a fixed universal simulator. On input \(\langle M\rangle\)
and \(x\), \(U(\langle M\rangle,x)\) simulates \(M(x)\). The existence
of a universal machine goes back to Turing's original construction
\cite{turing1937computable}; the modern textbook presentation is
standard \cite{hopcroft2001introduction}. Two clocks are distinguished.
The \emph{simulated clock} counts transitions of \(M\). The
\emph{decision clock} counts transitions of a machine attempting to
decide a bounded-halting predicate. They are distinguished for
accounting, but the diagonal arguments place both on the same fixed
unit-cost transition scale and compare them for a fixed budget \(T\).

\begin{definition}[Deterministic bounded halting]
\label{def:bh}
For a deterministic machine \(M\), input \(x\), and integer \(T>0\),
define \(BH_T(M,x)\) by
\[
\begin{aligned}
BH_T(M,x)=1
&\Longleftrightarrow M(x) \text{ accepts}\\
&\quad\text{within } T \text{ transitions.}
\end{aligned}
\]
Rejecting before \(T\), looping beyond \(T\), and not halting before
\(T\) all give value \(0\). The analogous halting version is equivalent
for the present purpose after the usual change of final states.
\end{definition}

\begin{definition}[Nondeterministic bounded halting]
\label{def:nbh}
For a nondeterministic machine \(N\), input \(x\), and integer \(T>0\),
define \(NBH_T(N,x)\) by
\[
\begin{aligned}
NBH_T(N,x)=1
&\Longleftrightarrow \text{some branch accepts}\\
&\quad\text{within depth } T.
\end{aligned}
\]
Thus \(T\) bounds the height of the computation tree. The definition is
the finite-budget version of the usual existential acceptance
semantics for nondeterministic time.
\end{definition}

\begin{definition}[Observable property]
\label{def:observable}
An observable property is a property inferred from the implementation
boundary (for example, memory size, transition budget, external
interfaces, or retention policy) without requiring a complete semantic
analysis of the model being implemented.
\end{definition}

Observable properties are intentionally coarse. They are useful
because implementation facts can refute universality even when the
internal model is complicated. A finite-state transducer may be large,
probabilistic, or difficult to inspect. If the whole closed system has a
finite set of configurations that is fixed independently of \(T\), then
it fails as a standalone universal machine for all budgets.

The definitions above also fix the role of the budget. In
Definitions~\ref{def:bh} and~\ref{def:nbh}, \(T\) denotes the number of
local transitions that the simulated run or branch is allowed to take.

Two kinds of decider for \(BH_T\) are compared below. A \emph{uniform}
decider is a single machine \(H(p,x,T)\) that takes a program code
\(p\), an input \(x\), and a budget \(T\), and returns \(BH_T(p,x)\),
the value of Definition~\ref{def:bh} for the machine coded by \(p\). A
\emph{fixed-budget} decider \(H_T(p, x)\) is specialized to one budget:
it returns \(BH_T(p,x)\) for that fixed \(T\), and \(\{H_T\}\) denotes
the resulting family. A uniform decider may be constrained by the need
to read an encoding of \(T\), but the lower bound below is stronger
because it is phrased for the fixed-budget family, thus removing the
possible artifact of reading a long unary clock from the input.

This convention is close to how resource-bounded verification is used.
A model checker, symbolic executor, or test harness is often asked to
search up to a specific depth
\cite{biere1999symbolic,king1976symbolic,cadar2013symbolic}.
The answer is expected for that chosen depth. Constants and lower-order terms
therefore matter operationally even when they would disappear in a
complexity-class statement.

\section{Related Work and Positioning}
\label{sec:related}

The paper uses standard notions from computability and automata theory:
Turing machines, universal simulation, finite automata, pushdown
automata, and nondeterministic acceptance
\cite{turing1937computable,hopcroft2001introduction}. The new emphasis
is the finite engineering reading placed on top of these standard
definitions.

\subsection{Bounded Halting and Asymptotic Hierarchies}

Bounded halting is a standard topic in complexity theory. The
time-hierarchy tradition begins with Hartmanis and Stearns
\cite{HartmanisStearns1965}. Trevisan's Berkeley notes use a universal
bounded-computation language \cite{Trevisan2008}:
\[
\begin{aligned}
U=\{(\langle M\rangle,x,t,\ell):{}&
\exists y,\ |y|\leq \ell,\\
&M(x,y)\text{ accepts within }t\text{ steps}\}.
\end{aligned}
\]
The integers \(t\) and \(\ell\) are encoded in unary in that
presentation. A nondeterministic verifier guesses \(y\) and checks the
\(t\)-step computation, so \(U\in NP\).

The analogous class-style statement for the present notation is the
language family
\[
L_p^{NBH}=
\{(\langle N\rangle,x): NBH_{p(n)}(N,x)=1\},
\]
where \(n=|(\langle N\rangle,x)|\). It belongs to
\(NTIME(O(p(n)))\), up to the fixed universal simulation overhead.
Alternatively, if \(T\) itself is supplied in unary, then
\[
\begin{aligned}
L^{NBH}_{unary}=\{(\langle N\rangle,x,1^T):{}&
NBH_T(N,x)=1\}.
\end{aligned}
\]
This is an \(NP\) language. A nondeterministic machine can verify a
yes-instance by following one branch of length \(p(n)\), while a
deterministic exhaustive simulator traverses the computation tree.
For a fixed model whose nondeterministic branching degree is bounded by
\(b\), this gives the usual contrast
\[
\begin{array}{rcl}
\mbox{branch verification} &:& NTIME(O(p(n))),\\
\mbox{tree traversal} &:& DTIME(b^{p(n)}p(n)^{O(1)}).
\end{array}
\]

This class comparison is useful for asymptotic separation. The
engineering contract studied here asks a sharper constant-level
question. Trevisan's notes use polynomial bounds such as \(|x|^3\) and
prove hierarchy statements in terms such as \(o(n^3)\). Such notation
is natural for separating complexity classes and leaves constants
outside the statement: a lower bound stated only against \(o(n^3)\)
still permits solving the same family in \(0.1\,n^3\) steps. That
speedup is still \(\Theta(n^3)\), hence asymptotically invisible, but it
can be decisive for an implementation. The present paper therefore
treats \(T\) as a hard budget.

The nondeterministic hierarchy also has a classical literature. Cook
proved a nondeterministic time hierarchy theorem, later sharpened by
Seiferas, Fischer, and Meyer \cite{Cook1973,SeiferasFM78}. The finite
reading below works alongside those results and isolates the
implementation question faced by a system that must return one stable
exact bit for the concrete predicate \(NBH_T\).

\subsection{Speedup and Fixed Implementations}

The present fixed-implementation result addresses a different question
from classical speedup theorems. Linear speedup reduces multiplicative
constants by changing the simulating machine, for instance by packing
several simulated tape symbols into one transition
\cite{HartmanisStearns1965}. Blum's
speedup theorem is different again: it exhibits functions with no
fastest program, so further speedups are obtained by changing the
program \cite{Blum1967}. Both statements are compatible with the present
claim. Here the implementation boundary, machine model, and exact
output contract are fixed first; the question is whether that fixed
artifact can decide every \(T\)-step bounded-halting instance in
\(\alpha T\) internal steps for some fixed \(\alpha<1\).

\subsection{Finitary and Infinitary Readings}

Each fixed budget \(BH_T\) is decidable by the bounded search
\(\exists\, t\le T\). The unbounded question has the form \(\exists\, t\)
with no finite bound and is only \emph{semidecidable}: acceptance is
eventually witnessed, but non-acceptance never certifies \(t<\infty\).
Classical undecidability of halting is exactly this gap, and it is
located only in the passage from a finite budget to no budget
\cite{tao2008structure}.

Two classical vocabularies name the same gap. In domain-theoretic
language, the unbounded predicate \(BH_\infty\) is the directed supremum
of the decidable approximations \(\{BH_T\}\) \cite{abramsky1994domain}:
it is the semidecidable \(\exists\, t\), with no decidable \(t<\infty\)
read off at one largest finite \(T\). Nonstandard analysis makes the same
cut with a hyperfinite budget \(T^{*}\): running for \(T^{*}\) steps is
an internally bounded search, so by transfer \(BH_{T^{*}}\) is decidable
\cite{robinson1966nonstandard}, and since any acceptance occurs at a
standard finite step \(t\le T^{*}\), it settles halting. The problem is
decidable in \(T^{*}\) but not in any standard finite \(T\).

\subsection{Machine-Learning Systems}

Machine-learning systems motivate the implementation-boundary
question. On the idealized side, attention models can be universal once
arbitrary numerical precision and unbounded intermediate computation
are assumed: P{\'e}rez et al.\ prove Turing-completeness for
hard-attention Transformers under such assumptions
\cite{perez2021attention}, and Merrill and Sabharwal show that the
number of generated chain-of-thought tokens is itself a computational
resource \cite{merrill2023expressive}. Formal Transformer limitations
are also known: self-attention is limited in language recognition unless
heads or layers grow with input length \cite{Hahn2020}; saturated
Transformers sit within constant-depth threshold circuits
\cite{MerrillSabharwalSmith2022}; and log-precision Transformers face
what Merrill and Sabharwal call a parallelism tradeoff: one forward
pass chains at most as many steps as its fixed depth, so a longer
sequential task needs generated tokens or repeated calls
\cite{MerrillSabharwal2023}. For recurrent models,
Weiss, Goldberg, and Yahav show that finite precision lowers
language-recognition power and blocks unbounded counting tasks that
arbitrary-precision idealizations can solve \cite{weiss2018practical}.

This paper uses those results to separate ideal model assumptions from
deployed finite-resource systems. A fixed finite-precision model with
fixed context and bounded runtime is analyzed here through the same
observable-resource tests as any other implementation.

\section{Deterministic Finite Reading}
\label{sec:deterministic}

The unbounded halting problem says that no total computable function
decides halting for all programs and inputs \cite{davis1958computability}.
A finite implementation changes this statement in a simple way.

\begin{proposition}[Fixed finite state]
\label{prop:finite-halting}
Consider a closed deterministic implementation whose set of reachable
configurations has size \(C\), where the same \(C\) is fixed
independently of the requested bounded-halting budget \(T\). For any
loaded request, its halting behavior is decidable by simulating at most
\(C+1\) transitions.
\end{proposition}

\begin{proof}
During a run, either the machine reaches a final configuration within
the first \(C+1\) transitions, or two configurations repeat. In the
second case determinism forces the future run to repeat the same cycle
forever. Hence the machine does not halt after the first repeated
configuration unless it already halted before the repetition.
\end{proof}

This proposition is sometimes used to dismiss the halting problem as
irrelevant to real machines. That conclusion is misleading. The
configuration bound \(C\) can be so large that enumeration has no
engineering value, and for the fixed machine that has actually been
built this is exactly the point: exact classification may be finite in
principle while still requiring the full price of search. The practical
question asks what the fixed implementation can decide within the stated
budget.

The bounded problem gives the practical replacement. For a concrete
budget \(T\), the trivial exact method is to simulate \(M(x)\) for
\(T\) steps and observe whether acceptance occurs. The following
diagonal theorem states the sense in which this method is worst-case
optimal for exact uniform decision. It is stated as a fixed-budget
family theorem to isolate the decision cost from the cost of reading the
encoding of \(T\).

\begin{theorem}[Deterministic diagonal bound]
\label{thm:det-bound}
Fix an effective coding of deterministic machines and a unit-cost
transition model. Let \(c\) be a constant large enough to cover the
following wrapper overhead: after receiving the answer of a candidate
decider, the wrapper can either enter a fixed nonaccepting infinite
loop or accept within at most \(c\) additional transitions.

For all sufficiently large \(T\), there is no deterministic machine
\(H_T\) such that, for every machine code \(p\) and input \(x\),
\[
H_T(p,x)=BH_T(p,x)
\]
and \(H_T\) halts with that answer in at most \(T-c\) transitions.
Consequently, no exact uniform bounded-halting decider can have
worst-case decision time below \(T-O(1)\) on all fixed-budget diagonal
instances.
\end{theorem}

\begin{proof}
Assume that such an \(H_T\) exists. Construct a deterministic machine
\(D_T\) which, on input \(z\), runs \(H_T(z,z)\). If the answer is
\(0\), then \(D_T\) accepts. If the answer is \(1\), then \(D_T\) enters
a fixed infinite loop. By the choice of \(c\), whenever \(H_T(z,z)\)
halts within \(T-c\) transitions and returns \(0\), the accepting run
of \(D_T(z)\) finishes within \(T\) transitions.

Let \(d\) be the code of \(D_T\) and run \(D_T(d)\). If
\(H_T(d,d)=1\), then by correctness \(D_T(d)\) accepts within \(T\)
transitions. But the definition of \(D_T\) makes it loop forever after
receiving the answer \(1\), a contradiction. If \(H_T(d,d)=0\), then \(D_T(d)\)
accepts within \(T\) transitions, so correctness requires
\(H_T(d,d)=1\), again a contradiction. Therefore no such \(H_T\)
exists.
\end{proof}

\begin{corollary}[No fixed-factor exact shortcut]
\label{cor:no-factor-shortcut}
Under the assumptions of Theorem~\ref{thm:det-bound}, fix any constant
\(\alpha<1\). For all sufficiently large \(T\), no exact deterministic
decider for \(BH_T\) can halt on all inputs within \(\alpha T\)
transitions.
\end{corollary}

\begin{proof}
Choose \(T\) large enough that \(\alpha T\leq T-c\). A decider running
within \(\alpha T\) transitions would then satisfy the forbidden
\(T-c\) bound of Theorem~\ref{thm:det-bound}.
\end{proof}

The bound is a worst-case, uniform statement, and many programs can
be classified statically or by short partial execution
\cite{cousot1977abstract,cook2011proving}. For any exact method that
handles all program-input pairs, there are diagonal instances that
require more than \(T-c\) transitions.
Corollary~\ref{cor:no-factor-shortcut} makes the practical consequence
explicit: after the machine model and wrapper overhead are fixed, even
a constant-factor exact shortcut such as \(T/2\) is impossible
uniformly for all sufficiently large budgets.

The theorem concerns exact decision; probabilistic and heuristic
predictions use a different contract. The constant \(c\) covers wrapper
steps after the decider returns and depends on the fixed machine model,
coding, and simulator. A universal accelerator may speed up many
programs, but the all-input contract includes the diagonal program.
The theorem gives a reason for using bounded halting as an
engineering universality certification test. A proposed universal accelerator can speed up
many useful programs. Under the exact all-input bounded-halting
contract, the diagonal program still forces the accelerator to meet the
bound it is asked to decide. The diagonal program is adversarial, but
the universality claim includes adversarial programs.

The usual infinite contradiction is recovered by letting the budget escape every
finite bound. The engineering reading keeps the finite number \(T\) in
view: a device that claims to decide all \(T\)-step behavior has to
spend, on some inputs, essentially the same order of resources as the
behavior being classified.

\section{The Budget as a Branching Limit}
\label{sec:budget}

The phrase ``bounded halting problem'' carries an engineering reading
that the asymptotic tradition obscures. In that tradition the bound is a
function of input length and the question is which class of functions
grows faster; that comparison, and how it relates to the present
predicates, is taken up in Section~\ref{sec:related}. In the
engineering reading the bound is a contract: the run is cut off after
\(T\) steps because a deadline, energy budget, recursion limit, or search
depth says so. A monitor with a budget of \(10^{9}\) transitions must
decide whether the watched run accepts by that budget. The number
\(10^{9}\) is the actual limit. The hard-budget reading therefore asks the
sharper question: can an exact decider uniformly answer a \(T\)-step
question in \(\alpha T\) steps for some fixed \(\alpha<1\)? The diagonal
bound says no, once the model overhead has been fixed.

For deterministic machines the finite picture is a path of
configurations
\[
C_0 \to C_1 \to \cdots \to C_T.
\]
Deciding \(BH_T\) asks whether an accepting configuration appears on
that prefix. The direct algorithm walks the prefix. A faster exact
algorithm must somehow infer a property of the unseen suffix of the
prefix. The diagonal theorem shows why this cannot be guaranteed for
all programs: the program being analyzed can use the analyzer's early
answer to make the answer false.

For nondeterministic machines the finite picture is a rooted tree. The
root is the initial configuration, and every node has one child for
each legal nondeterministic transition. The predicate \(NBH_T\) asks
whether there is an accepting node at depth at most \(T\). The width of
the tree may be exponential in \(T\), but the branch-depth budget is
still a sequential resource along every certificate path. A certificate
for acceptance is a path, and the path contains at most \(T\) local
transitions that must be checked.

\section{Nondeterministic Finite Reading}
\label{sec:nondeterministic}

Nondeterminism is often described as parallel exploration of many
possible computations. The bounded predicate \(NBH_T\) makes the
resource interpretation precise. The budget \(T\) is the maximum branch
depth, while the total number of nodes in the tree may be exponential in
\(T\). A branch may be guessed, but an accepting branch of length \(T\)
still contains \(T\) successive transitions.

A nondeterministic universal simulator decides \(NBH_T(N,x)\) by
guessing one branch and simulating at most \(T\) transitions, up to the
fixed overhead of universal simulation. This section asks the matching
lower-bound question for a fixed budget.

The nondeterministic lower-bound situation is more delicate than the
deterministic one. For ordinary nondeterministic acceptance 
one accepting branch is enough. Negating such a statement asks for a
universal condition over all branches, which is why nondeterministic
hierarchy theorems require more subtle arguments than the elementary
deterministic diagonal proof. The hierarchy background is reviewed in
Section~\ref{sec:related}. From the engineering standpoint, branching
changes width, while the finite reading of \(T\) as depth remains.

For systems that use nondeterminism as an actual decision procedure, the
specification also states how an observer obtains a single answer. The
next statement records the strong engineering form.

\begin{theorem}[Stable nondeterministic bound]
\label{thm:nondet-stable}
Fix the model and overhead constant \(c\) as in
Theorem~\ref{thm:det-bound}. There is no nondeterministic family
\(A_T\) that exactly decides \(NBH_T\) in at most \(T-c\) transitions
on every branch and has stable observable output, meaning that all
terminating branches return the same bit for each input.
\end{theorem}

\begin{proof}
Suppose such a family \(A_T\) exists. Restrict its inputs to
deterministic machines. For deterministic \(M\), the predicate
\(NBH_T(M,x)\) is identical to \(BH_T(M,x)\), because the computation
tree has one path. Since \(A_T\) has stable output, a deterministic
machine can follow any fixed legal branch of \(A_T\) and obtain the
same answer within the same time bound. This gives a deterministic
family deciding \(BH_T\) within \(T-c\), contradicting
Theorem~\ref{thm:det-bound}.
\end{proof}

The same reduction against Corollary~\ref{cor:no-factor-shortcut}
sharpens this to a constant factor: no stable nondeterministic decider
settles \(NBH_T\) within \(\alpha T\) on every branch for a fixed
\(\alpha<1\). This answers the question left open in
Section~\ref{sec:related}: the branch-depth budget cannot be beaten,
even by a constant factor.

Without stability, a recognizer may accept on one branch and provide no
one-path witness for a negative answer. With stability, the
deterministic obstruction applies to deterministic subinstances. Such an
instance has one path, and transition \(t+1\) depends on transition
\(t\). Parallel speculation becomes exact only after the true
continuation is verified. Branching can reduce search width but cannot
remove this causal depth.

\begin{corollary}[Probability-one exactness]
\label{cor:prob-one}
No finite randomized implementation whose possible branches all have
positive probability can decide \(NBH_T\) with probability \(1\) and
halt on every branch within \(T-c\) transitions.
\end{corollary}

\begin{proof}
Probability-one correctness implies that no positive-probability branch
returns a wrong bit. Thus all branches return the same exact bit, giving
the stable decider prohibited by Theorem~\ref{thm:nondet-stable}.
\end{proof}

Random sampling may instead provide a confidence level, a failure
probability, or an ``unknown'' outcome. Probability-one exactness in the
finite model retains the diagonal bound.

\section{Observable Properties of Nonuniversal Machines}
\label{sec:observable}

The previous sections use bounded halting to compare a decision
procedure with the computation it decides. The direction is now
reversed. For a universal machine, the bounded halting problem cannot be
decided faster than simulation. Given an implementation, what can be
inferred about its universality by using bounded halting as a test,
without understanding all its internal semantics?

The following properties are observable in the sense of
Definition~\ref{def:observable}. They are practical refutation tests:
if an implementation has one of them as a standalone machine, then it
cannot be a universal machine for exact all-budget computation. Passing
the tests leaves the constructive simulation proof as the next step.

Addressable memory, update rounds, context length, arithmetic precision,
and persistent storage can often be determined from the implementation
boundary.

\subsection{Configuration Space Independent of \(T\)}

For every finite \(T\), a Turing-machine run visits finitely many
configurations. The obstruction is one finite configuration set fixed
before \(T\) is chosen. This differs from a fixed computation model that
permits its tape or stack to grow as \(T\) grows.

If the whole closed implementation has such a fixed finite number of
configurations after input loading, then its behavior is ultimately
periodic by Proposition~\ref{prop:finite-halting}. This includes
ordinary finite-state machines and also very large finite controllers
with bounded memory. Classification depends on fixedness. A machine
with \(2^{10^{12}}\) possible states is still finite if the same state
set is used for every requested \(T\); the changing work space required
to pay for arbitrary larger searches is absent.

A system may include neural components, probabilistic choices, caches,
and schedulers. If all are bounded independently of \(T\), the global
state graph remains finite. Randomness gives a finite probabilistic
transition system with the same boundary.

\subsection{Bounded Deliberation}

A universal machine must be able to spend more time when the simulated
program requires more time. Therefore a standalone implementation with
a fixed response bound cannot be universal for unbounded computation.

\begin{proposition}[Fixed deliberation]
\label{prop:deliberation}
Suppose an implementation always produces its final answer after at
most \(B\) internal transitions, where \(B\) is independent of the
program being simulated and of the requested budget \(T\). Then the
implementation cannot be an exact universal bounded-halting decider for
all \(T\).
\end{proposition}

\begin{proof}
Choose \(T>B+c\), where \(c\) is the fixed wrapper overhead from
Theorem~\ref{thm:det-bound}. If the implementation exactly decided
\(BH_T\) for all program-input pairs within \(B\) transitions, it would
decide \(BH_T\) within \(T-c\) transitions. This contradicts
Theorem~\ref{thm:det-bound}.
\end{proof}

A feed-forward circuit, fixed-depth inference graph, or controller with
a fixed number of update rounds remains a bounded transducer. A larger
fixed \(B\) handles more bounded tasks but does not supply an unbounded
time parameter.

This test observes only the absence of a fixed deliberation bound. It
does not establish a chain of useful dependent computations. A machine
can pass the test by repeating one state forever. Resource growth and
accessible intermediate data are separate requirements.

\subsection{Inaccessible Intermediate Data}

The next simple observable property is memory retention. A universal
machine is programmable so that data produced at one time can affect
computation arbitrarily far in the future. A program may discard intermediate
symbols, and the model allows it to preserve or reconstruct data when
the program requires it.

\begin{proposition}[Finite forgetting]
\label{prop:forgetting}
Consider a sequential implementation with a fixed finite accessible
state space \(S\), where \(S\) is independent of \(T\). 
If, after some fixed horizon \(h\) independent of \(T\), earlier input or
intermediate data can affect future outputs only through the current
element of \(S\), then the implementation cannot realize all
computations of a universal machine.
\end{proposition}

\begin{proof}
Choose \(n\geq h\) such that \(2^n>|S|\). By the pigeonhole principle,
two distinct binary strings \(u,v\in\{0,1\}^n\) leave the implementation
in the same accessible state. Give both runs the same continuation
\(\#u\). Their future behavior must be identical because earlier input
can affect it only through the current element of \(S\). Let \(M_{\rm
copy}\) be a Turing machine deciding
\(\{w\#w:w\in\{0,1\}^*\}\), and choose \(T\) larger than its running
time on these two inputs. Then
\[
\begin{aligned}
 BH_T(M_{\rm copy},u\#u)&=1,\\
 BH_T(M_{\rm copy},v\#u)&=0.
\end{aligned}
\]
The implementation gives the same answer on both inputs and therefore
fails at least one of these bounded-halting instances. Thus it cannot realize every
Turing-machine computation.
\end{proof}

Forgetting is often appropriate: many controllers
discard old data for stability, privacy, or bounded resource use. What
the proposition isolates is that irreversible finite forgetting is an
observable obstruction to standalone universality.

The same test applies to intermediate data as well as input data.
Suppose a computation derives a lemma, a partial parse, a stack frame,
or a simulated tape segment and later overwrites every accessible trace
of it. If the implementation has no unbounded store in which a program
could choose to save that trace, then later behavior cannot depend on
the lost distinction. A universal machine allows the opposite programming
pattern: compute a value now, keep it, and use it after an arbitrarily
long delay.

\subsection{An Audit Checklist}

The audit separates the mathematical universality of the parametric model
from the resources used by a fixed implementation. Four questions guide the 
audit before a simulation proof.

\begin{enumerate}
\item \emph{Boundary.} What exactly is the machine: a neural network, an
interpreter process, an operating system with files, or a
human-in-the-loop workflow? A component and its surrounding system have
different boundaries. A CPU may access extendable memory, and an agent
may append to external files (see CSP~\cite{hoare1985communicating}).
Tape or stack growth extends the resources of a fixed computation model;
changes in precision, parameter count, architecture, or exactly
addressable context define a growing model family.

\item \emph{Accounting.} Where is the cost of exact search paid? If no
resource can grow with \(T\), the closed system is
finite-state~(Proposition~\ref{prop:finite-halting}).

\item \emph{Clock.} What allows the machine to continue, and is
that continuation part of the model? A component with a hard maximum
number of rounds fails Proposition~\ref{prop:deliberation}.

\item \emph{Retention.} Where can a program place data that must be used
later? A fixed cache, context window, or hidden state can summarize old
information. Persistent files, logs, stacks, or databases can preserve a
growing history when their capacity is unbounded and the program can
access them~(Proposition~\ref{prop:forgetting}).
\end{enumerate}

\section{Application to Machine Classes}
\label{sec:applications}

Table~\ref{tab:machine-classes} applies the three observable tests:
configuration growth with \(T\), unbounded deliberation, and accessible
intermediate data. Passing all three still requires a simulation proof.

\begin{table*}[t]
\centering
\footnotesize
\caption{Yes/no application of the three observable tests}
\label{tab:machine-classes}
\setlength{\tabcolsep}{6pt}
\begin{tabular}{p{0.44\textwidth}cccc}
\hline
Class & Growth & Deliberation & Later data & All tests\\
\hline\hline
Finite-state machine \cite{hopcroft2001introduction} & No & No & No & No\\
\hline
\(\varepsilon\)-transition finite-state machine
\cite{hopcroft2001introduction} & No & Yes & No & No\\
\hline
Pushdown automaton \cite{hopcroft2001introduction} & Yes & Yes & No & No\\
\hline
Two-stack pushdown automaton \cite{hopcroft2001introduction} & Yes & Yes & Yes & Yes\\
\hline
Unbounded CSP-style multi-agent system
\cite{hoare1985communicating} & Yes & Yes & Yes & Yes\\
\hline
Fixed finite-precision Transformer \cite{vaswani2017attention} & No & No & No & No\\
\hline
Fixed-context Transformer with unbounded generation
\cite{wei2022chain,merrill2023expressive} & No & Yes & No & No\\
\hline
\end{tabular}
\end{table*}

In Table~\ref{tab:machine-classes}, ``Growth'' means that the exact
configuration capacity available to the computation can grow with the
budget \(T\). The ``Deliberation'' column asks whether the system can
make a run with \(T\) internal transitions before producing its final
answer. It does not require progress or \(T\) useful dependent steps.
An \(\varepsilon\)-loop or a fixed-context generation loop passes this
column by repeating states. The ``Later data'' column asks whether
intermediate data can be retained and addressed later in the
computation. The last column is the conjunction of the three tests.

An \(\varepsilon\)-loop gives an FSM unbounded deliberation by repeating
a state, while configuration growth and addressable history remain
absent. It changes operational behavior through divergence but adds no
new storage; under ordinary acceptance semantics it is summarized by
\(\varepsilon\)-closure. Thus neither ordinary nor
\(\varepsilon\)-transition FSMs pass all three tests.

An unbounded pushdown stack supplies growth and deliberation but only
LIFO access: popped data is unavailable after removal. 
With two stacks, one can encode the left and right
sides of a Turing tape, giving the usual universal construction
\cite{hopcroft2001introduction}. 

Unbounded CSP-style multi-agent systems \cite{hoare1985communicating}
pass all three tests at the class level. The system boundary permits
growth in processes, communication, storage, or time. Persistent
messages or stores can retain intermediate data. Two unbounded stack
processes coordinated by a finite controller can simulate a Turing
machine, with storage and execution time growing as needed.
Fixed agents, memories, transcripts, and rounds instead define a finite
transition system.

The computational model of the multiagent real-time system
in \cite{krokhalev2026consensus} resembles CSP. If such a system is deployed with a
fixed response deadline that does not increase with the agent population,
its fixed computation and communication rates bound the work available per request. 
Under this deadline, it fails the fixed-deliberation test
(Proposition~\ref{prop:deliberation}), despite its distributed architecture.

The Transformer rows separate fixed forward computation from unbounded
generation \cite{vaswani2017attention}. Generated intermediate tokens
are a computational resource
\cite{wei2022chain,merrill2023expressive}, but unbounded generation
changes only deliberation. With fixed precision and context, a
deterministic decoder eventually repeats an accessible configuration
and loops; stochastic decoding remains finite-state and gives no exact
progress guarantee.

Transcript length supplies growth only when the computation can address
it exactly. Text outside fixed context is not component memory, and
positional information is limited by positional accuracy. Once more
positions are required than the implementation can distinguish and
address exactly, old data cannot act as arbitrary tape cells. A larger
system may retain the transcript externally, but certification then
applies to that combined boundary.

\section{Discussion}
\label{sec:discussion}

\subsection{Practical Universality}
A fixed language model has bounded parameters, precision, context,
sampling state, and internal work. Universality results instead assume
growth at the model boundary. Removing the generation-length bound does
not supply addressable growth: fixed-context generation can repeat states
and loop. The growth rule, not the physical finiteness of one artifact,
determines which model is being certified.

An external loop may add exactly addressable transcripts, files, tools,
program execution, or unbounded scheduler time. Certification then
applies to the combined system, provided that later computation can
access these resources. An unbounded CSP-style system
(Table~\ref{tab:machine-classes}) or a general-purpose interpreter may
therefore be universal while each fixed component is not.

\subsection{Limitations of the Reading}

The result concerns exact answers for all program codes. Restricted
classes, sufficient conditions, ``unknown'' outputs, distributions, and
failure probabilities use different contracts and may permit faster
tools. Classical speedup results also change the machine, program, or
contract; here the implementation boundary and exact output remain
fixed.

\section{Conclusion}
\label{sec:conclusion}

The engineering question is whether a system with a stated boundary,
memory, precision, and clock can return exact answers for all
computations up to a budget \(T\). Bounded halting gives a minimal form of
this question. Some adversarial instances require exact simulation
through \(T\) transitions. An exact checker must handle these instances
within its stated resource contract.

Bounded halting also gives adversarial tests for universality
certification. These tests check which resources grow with \(T\),
whether internal deliberation has a fixed bound, and whether
intermediate data remain available. Unbounded deliberation alone can
be a loop and does not certify useful progress. The resource growth may
come from extendable tape or stack in a fixed computation model, from a
growing family of models, or from a larger system around a fixed
component. A universality claim must state this boundary, the growth
rule, the continuation rule, and the storage used by later steps.

\bibliography{main}

@book{hopcroft2001introduction,
  author    = {Hopcroft, John E. and Motwani, Rajeev and Ullman, Jeffrey D.},
  title     = {Introduction to Automata Theory, Languages, and Computation},
  edition   = {2nd},
  publisher = {Addison-Wesley},
  address   = {Boston},
  year      = {2001},
  comment   = {Summary: Standard textbook treatment of finite automata, pushdown automata, Turing machines, and universal simulation. Relevance: Supplies the conventional machine-class definitions used by the preliminaries and application table.}
}

@article{Blum1967,
  author  = {Blum, Manuel},
  title   = {A Machine-Independent Theory of the Complexity of Recursive Functions},
  journal = {Journal of the ACM},
  volume  = {14},
  number  = {2},
  pages   = {322--336},
  year    = {1967},
  doi     = {https://doi.org/10.1145/321386.321395},
  comment = {Summary: Introduces machine-independent complexity measures and the speedup theorem, exhibiting recursive functions with no fastest program. Relevance: One of the classical speedup results the fixed-artifact no-shortcut reading must coexist with.}
}

@article{turing1937computable,
  author  = {Turing, Alan M.},
  title   = {On Computable Numbers, with an Application to the {Entscheidungsproblem}},
  journal = {Proceedings of the London Mathematical Society},
  series  = {2},
  volume  = {42},
  number  = {1},
  pages   = {230--265},
  year    = {1937},
  doi     = {https://doi.org/10.1112/plms/s2-42.1.230},
  comment = {Summary: Introduces Turing machines and the universal-machine construction for computability. Relevance: Provides the classical foundation for the paper's universal-machine and halting-problem discussion.}
}

@book{davis1958computability,
  author    = {Davis, Martin},
  title     = {Computability and Unsolvability},
  publisher = {McGraw-Hill},
  address   = {New York},
  year      = {1958},
  comment   = {Summary: Textbook development, from lecture courses, that gives the standard formulation and proof of the unsolvability of the halting problem. Relevance: Provides the explicit attribution for the undecidability of the halting problem, which Turing's 1937 paper does not state in this form.}
}

@article{HartmanisStearns1965,
  author  = {Hartmanis, Juris and Stearns, Richard E.},
  title   = {On the Computational Complexity of Algorithms},
  journal = {Transactions of the American Mathematical Society},
  volume  = {117},
  pages   = {285--306},
  year    = {1965},
  doi     = {https://doi.org/10.1090/s0002-9947-1965-0170805-7},
  comment = {Summary: Establishes foundational time-complexity notions and hierarchy-style separation arguments. Relevance: Supports the deterministic bounded-halting and no-speedup reading.}
}

@article{Cook1973,
  author  = {Cook, Stephen A.},
  title   = {A Hierarchy for Nondeterministic Time Complexity},
  journal = {Journal of Computer and System Sciences},
  volume  = {7},
  number  = {4},
  pages   = {343--353},
  year    = {1973},
  doi     = {https://doi.org/10.1016/S0022-0000(73)80028-5},
  comment = {Summary: Gives an early hierarchy theorem for nondeterministic time. Relevance: Anchors the nondeterministic bounded-halting section and its branch-depth interpretation.}
}

@article{SeiferasFM78,
  author  = {Seiferas, Joel I. and Fischer, Michael J. and Meyer, Albert R.},
  title   = {Separating Nondeterministic Time Complexity Classes},
  journal = {Journal of the ACM},
  volume  = {25},
  number  = {1},
  pages   = {146--167},
  year    = {1978},
  doi     = {https://doi.org/10.1145/322047.322061},
  comment = {Summary: Proves stronger separations for nondeterministic time complexity classes. Relevance: Provides the standard hierarchy background for exact nondeterministic resource bounds.}
}

@misc{Trevisan2008,
  author       = {Trevisan, Luca},
  title        = {Notes for Lecture 1: {U.C. Berkeley CS278} Computational Complexity},
  howpublished = {Lecture notes},
  year         = {2008},
  note         = {Available at \url{https://lucatrevisan.github.io/cs278-08/lecture01.pdf}},
  comment      = {Summary: Introduces computational complexity, NP, bounded halting, and diagonalization in an asymptotic teaching frame. Relevance: Provides the lecture-note reading contrasted with the paper's finite-budget interpretation.}
}

@inproceedings{vaswani2017attention,
  author    = {Vaswani, Ashish and Shazeer, Noam and Parmar, Niki and Uszkoreit, Jakob and Jones, Llion and Gomez, Aidan N. and Kaiser, Lukasz and Polosukhin, Illia},
  title     = {Attention Is All You Need},
  booktitle = {Advances in Neural Information Processing Systems},
  volume    = {30},
  publisher = {Curran Associates, Inc.},
  year      = {2017},
  eprint    = {1706.03762},
  archivePrefix = {arXiv},
  primaryClass  = {cs.CL},
  comment   = {Summary: Introduces the Transformer architecture based on attention mechanisms. Relevance: Supplies the minimal transformer context for the finite-resource model discussion.}
}

@inproceedings{lewis2020retrieval,
  author        = {Lewis, Patrick and Perez, Ethan and Piktus, Aleksandra and Petroni, Fabio and Karpukhin, Vladimir and Goyal, Naman and K{\"u}ttler, Heinrich and Lewis, Mike and Yih, Wen-tau and Rockt{\"a}schel, Tim and Riedel, Sebastian and Kiela, Douwe},
  title         = {Retrieval-Augmented Generation for Knowledge-Intensive {NLP} Tasks},
  booktitle     = {Advances in Neural Information Processing Systems},
  volume        = {33},
  publisher     = {Curran Associates, Inc.},
  year          = {2020},
  eprint        = {2005.11401},
  archivePrefix = {arXiv},
  primaryClass  = {cs.CL},
  comment       = {Summary: Combines a sequence model with an external non-parametric memory accessed by retrieval. Relevance: Provides an example in which resources outside a fixed language-model component form part of the certified system boundary.}
}

@inproceedings{wu2024autogen,
  author        = {Wu, Qingyun and Bansal, Gagan and Zhang, Jieyu and Wu, Yiran and Li, Beibin and Zhu, Erkang and Jiang, Li and Zhang, Xiaoyun and Zhang, Shaokun and Liu, Jiale and Awadallah, Ahmed Hassan and White, Ryen W. and Burger, Doug and Wang, Chi},
  title         = {{AutoGen}: Enabling Next-Gen {LLM} Applications via Multi-Agent Conversations},
  booktitle     = {Proceedings of the First Conference on Language Modeling},
  year          = {2024},
  eprint        = {2308.08155},
  archivePrefix = {arXiv},
  primaryClass  = {cs.AI},
  comment       = {Summary: Builds language-model applications by composing multiple communicating agents with tools and human inputs. Relevance: Provides an example of a larger multi-agent system built around language-model components.}
}

@article{Hahn2020,
  author  = {Hahn, Michael},
  title   = {Theoretical Limitations of Self-Attention in Neural Sequence Models},
  journal = {Transactions of the Association for Computational Linguistics},
  volume  = {8},
  pages   = {156--171},
  year    = {2020},
  doi     = {https://doi.org/10.1162/tacl_a_00306},
  comment = {Summary: Proves formal-language limitations of self-attention models unless architectural resources such as layers or heads grow with input length. Relevance: Supports the paper's claim that fixed Transformer-style components have computational limitations under finite resource assumptions.}
}

@article{MerrillSabharwalSmith2022,
  author  = {Merrill, William and Sabharwal, Ashish and Smith, Noah A.},
  title   = {Saturated {Transformers} Are Constant-Depth Threshold Circuits},
  journal = {Transactions of the Association for Computational Linguistics},
  volume  = {10},
  pages   = {843--856},
  year    = {2022},
  doi     = {https://doi.org/10.1162/tacl_a_00493},
  comment = {Summary: Gives a constant-depth threshold-circuit upper bound for saturated Transformer variants. Relevance: Provides a circuit-complexity limitation result for Transformer-style models in the machine-learning systems discussion.}
}

@article{MerrillSabharwal2023,
  author  = {Merrill, William and Sabharwal, Ashish},
  title   = {The Parallelism Tradeoff: Limitations of {Log-Precision Transformers}},
  journal = {Transactions of the Association for Computational Linguistics},
  volume  = {11},
  pages   = {531--545},
  year    = {2023},
  doi     = {https://doi.org/10.1162/tacl_a_00562},
  comment = {Summary: Shows that log-precision Transformers admit highly parallel circuit simulations and therefore inherit corresponding computational limitations. Relevance: Supports the paper's distinction between practical parallel Transformer computation and exact universal simulation.}
}

@misc{wei2022chain,
  author        = {Wei, Jason and Wang, Xuezhi and Schuurmans, Dale and Bosma, Maarten and Ichter, Brian and Xia, Fei and Chi, Ed H. and Le, Quoc V. and Zhou, Denny},
  title         = {{Chain-of-Thought} Prompting Elicits Reasoning in Large Language Models},
  howpublished  = {arXiv preprint arXiv:2201.11903},
  year          = {2022},
  eprint        = {2201.11903},
  archivePrefix = {arXiv},
  primaryClass  = {cs.CL},
  comment       = {Summary: Defines chain-of-thought prompting as generating intermediate reasoning steps before an answer. Relevance: Justifies the paper's use of chain-of-thought as an unbounded generated transcript rather than a formal automata model.}
}

@misc{merrill2023expressive,
  author        = {Merrill, William and Sabharwal, Ashish},
  title         = {The Expressive Power of {Transformers} with {Chain of Thought}},
  howpublished  = {arXiv preprint arXiv:2310.07923},
  year          = {2023},
  eprint        = {2310.07923},
  archivePrefix = {arXiv},
  primaryClass  = {cs.CL},
  comment       = {Summary: Studies how the number of intermediate decoding steps changes the computational power of Transformer decoders. Relevance: Provides the main formal contrast for the paper's fixed-context and unbounded-chain Transformer row.}
}

@article{perez2021attention,
  author  = {P{\'e}rez, Jorge and Barcel{\'o}, Pablo and Marinkovic, Javier},
  title   = {Attention Is {Turing}-Complete},
  journal = {Journal of Machine Learning Research},
  volume  = {22},
  number  = {75},
  pages   = {1--35},
  year    = {2021},
  comment = {Summary: Shows Turing-completeness for idealized hard-attention Transformer models under mathematical resource assumptions. Relevance: Separates idealized attention-model universality from finite-resource implementations. URL: https://jmlr.org/papers/v22/20-302.html.}
}

@misc{weiss2018practical,
  author        = {Weiss, Gail and Goldberg, Yoav and Yahav, Eran},
  title         = {On the Practical Computational Power of Finite Precision {RNNs} for Language Recognition},
  howpublished  = {arXiv preprint arXiv:1805.04908},
  year          = {2018},
  eprint        = {1805.04908},
  archivePrefix = {arXiv},
  primaryClass  = {cs.CL},
  comment       = {Summary: Analyzes how finite precision and bounded recurrent computation change the formal language-recognition power of neural sequence models. Relevance: Supports the paper's distinction between ideal neural-model universality and finite-resource implementations.}
}

@book{hoare1985communicating,
  author    = {Hoare, C. A. R.},
  title     = {Communicating Sequential Processes},
  publisher = {Prentice Hall},
  address   = {Englewood Cliffs, NJ},
  year      = {1985},
  comment   = {Summary: Presents the CSP process formalism for communicating concurrent systems. Relevance: Supports the application-table row treating unbounded multi-agent process systems.}
}

@book{tao2008structure,
  author    = {Tao, Terence},
  title     = {Structure and Randomness: Pages from Year One of a Mathematical Blog},
  publisher = {American Mathematical Society},
  address   = {Providence, RI},
  year      = {2008},
  comment   = {Summary: Collects expository essays, including ``Soft analysis, hard analysis, and the finite convergence principle'' on the finitary-infinitary correspondence. Relevance: Places the paper's finite-budget reading within the general finitary-infinitary correspondence.}
}

@book{robinson1966nonstandard,
  author    = {Robinson, Abraham},
  title     = {Non-standard Analysis},
  publisher = {North-Holland},
  address   = {Amsterdam},
  year      = {1966},
  comment   = {Summary: Founds non-standard analysis through the hyperreals and hypernaturals, internal sets, the transfer principle, and the standard-part map. Relevance: Supplies the hyperfinite-budget reading of bounded halting in the finitary-infinitary discussion.}
}

@inproceedings{biere1999symbolic,
  author    = {Biere, Armin and Cimatti, Alessandro and Clarke, Edmund M. and Zhu, Yunshan},
  title     = {Symbolic Model Checking without {BDDs}},
  booktitle = {Tools and Algorithms for the Construction and Analysis of Systems (TACAS)},
  series    = {Lecture Notes in Computer Science},
  volume    = {1579},
  pages     = {193--207},
  publisher = {Springer},
  address   = {Berlin, Heidelberg},
  year      = {1999},
  doi       = {https://doi.org/10.1007/3-540-49059-0_14},
  comment   = {Summary: Introduces bounded model checking, which searches for counterexamples up to a fixed unrolling depth k. Relevance: Concrete instance of the depth-bounded verification reading motivating the fixed-budget contract.}
}

@article{king1976symbolic,
  author  = {King, James C.},
  title   = {Symbolic Execution and Program Testing},
  journal = {Communications of the ACM},
  volume  = {19},
  number  = {7},
  pages   = {385--394},
  year    = {1976},
  doi     = {https://doi.org/10.1145/360248.360252},
  comment = {Summary: Founds symbolic execution as a method that explores program paths, in practice up to a bounded depth. Relevance: Supports the depth-bounded search reading of symbolic executors and test harnesses.}
}

@article{cadar2013symbolic,
  author  = {Cadar, Cristian and Sen, Koushik},
  title   = {Symbolic Execution for Software Testing: Three Decades Later},
  journal = {Communications of the ACM},
  volume  = {56},
  number  = {2},
  pages   = {82--90},
  year    = {2013},
  doi     = {https://doi.org/10.1145/2408776.2408795},
  comment = {Summary: Surveys modern symbolic-execution test-generation tools, which bound path-exploration depth to control search cost. Relevance: Connects the depth-bounded reading to current test-harness practice.}
}

@inproceedings{cousot1977abstract,
  author    = {Cousot, Patrick and Cousot, Radhia},
  title     = {Abstract Interpretation: A Unified Lattice Model for Static Analysis of Programs by Construction or Approximation of Fixpoints},
  booktitle = {Proceedings of the 4th ACM SIGACT-SIGPLAN Symposium on Principles of Programming Languages (POPL)},
  pages     = {238--252},
  publisher = {ACM},
  address   = {New York},
  year      = {1977},
  doi       = {https://doi.org/10.1145/512950.512973},
  comment   = {Summary: Founds abstract interpretation, the framework for sound static classification of program behavior. Relevance: Supports the claim that many programs can be classified statically without full simulation.}
}

@article{cook2011proving,
  author  = {Cook, Byron and Podelski, Andreas and Rybalchenko, Andrey},
  title   = {Proving Program Termination},
  journal = {Communications of the ACM},
  volume  = {54},
  number  = {5},
  pages   = {88--98},
  year    = {2011},
  doi     = {https://doi.org/10.1145/1941487.1941509},
  comment = {Summary: Surveys automated termination analyses that settle halting for large classes of real programs. Relevance: Supports the claim that many programs are classifiable in practice even though the general problem is hard.}
}

@article{krokhalev2026consensus,
  author  = {Krokhalev, Evgeny and Arkhipov, Ivan and Granichin, Oleg and Tarasova, Elizaveta},
  title   = {Consensus Algorithms for a Multiagent Operating System},
  journal = {Cybernetics and Physics},
  volume  = {15},
  number  = {1},
  pages   = {38--55},
  year    = {2026},
  doi     = {https://doi.org/10.35470/2226-4116-2026-15-1-38-55},
  comment = {Summary: Studies logical-time synchronization and cooperative tracking in a multiagent operating system with communication delays. Relevance: Provides an engineering example for distinguishing extensible CSP-style computation from a fixed implementation subject to a response deadline; the deadline is a deployment assumption made in the present paper.}
}

@incollection{abramsky1994domain,
  author    = {Abramsky, Samson and Jung, Achim},
  title     = {Domain Theory},
  booktitle = {Handbook of Logic in Computer Science},
  editor    = {Abramsky, S. and Gabbay, D. M. and Maibaum, T. S. E.},
  volume    = {3},
  pages     = {1--168},
  publisher = {Oxford University Press},
  address   = {Oxford},
  year      = {1994},
  comment   = {Summary: Standard reference on dcpos, Scott continuity, and computation as directed suprema of finite approximations. Relevance: Supplies the domain-theoretic reading of the infinite-budget predicate as a directed supremum of finite-budget approximations.}
}

\end{document}